\documentclass[useAMS,usenatbib]{mnras}
\usepackage{graphicx}
\usepackage{textcomp}
\usepackage{amssymb}\usepackage{hyperref}
\usepackage{amsmath,alltt}                                                                                                                     
\usepackage{multirow}
\usepackage{rotating}
\usepackage{lscape}
 \usepackage{times}

\newcommand\lsim{\mathrel{\rlap{\lower4pt\hbox{\hskip1pt$\sim$}}
        \raise1pt\hbox{$<$}}}
\newcommand\gsim{\mathrel{\rlap{\lower4pt\hbox{\hskip1pt$\sim$}}
        \raise1pt\hbox{$>$}}}
\newcommand{\be}{\begin{equation}}
\newcommand{\ba}{\begin{eqnarray}}
\newcommand{\ee}{\end{equation}}
\newcommand{\ea}{\end{eqnarray}}

\title[On the rate of BH binary mergers in galactic nuclei]{On the rate of black hole binary mergers in galactic nuclei due to dynamical hardening}

\author[N. W. C. Leigh et al.]{N. W. C. Leigh$^{1,2,3}$, A. M. Geller$^{4,5}$, B. McKernan$^{1,6,7}$, K.E.S. Ford$^{1,6,7}$, 
\newauthor
M.-M. Mac Low$^{1}$, J. Bellovary$^{1,8}$, Z. Haiman$^{9,10}$, W. Lyra$^{11}$,  
\newauthor
J. Samsing$^{12}$, M. O'Dowd$^{1,6,13}$, B. Kocsis$^{14}$, S. Endlich$^{15}$\\
$^{1}$Department of Astrophysics, American Museum of Natural History, New York, NY 10024, USA\\
$^{2}$Department of Physics and Astronomy, Stony Brook University, Stony Brook, NY 11794-3800, USA\\
$^{3}$Center for Computational Astrophysics, Flatiron Institute, 162 Fifth Avenue, New York, NY 10010, USA\\
$^{4}$Center for Interdisciplinary Exploration and Research in Astrophysics (CIERA) and Department of Physics and Astronomy, \\ Northwestern University, 2145 Sheridan Rd, Evanston, IL 60208, USA \\
$^{5}$Adler Planetarium, Dept.\ of Astronomy, 1300 S. Lake Shore Drive, Chicago, IL 60605, USA \\
$^{6}$Graduate Center, City University of New York, 365 5th Avenue, New York, NY 10016, USA\\
$^{7}$Department of Science, Borough of Manhattan Community College, City University of New York, New York, NY 10007, USA\\
$^{8}$Department of Physics, Queensborough Community College, City University of New York, Bayside, NY 11364, USA \\
$^{9}$Department of Astronomy, Columbia University, New York, NY, USA\\
$^{10}$Center for Cosmology and Particle Physics, New York University, New York, NY, USA\\
$^{11}$Department of Physics \& Astronomy, California State University Northridge, 18111 Nordhoff St., Northridge CA 91330, USA \\
$^{12}$Department of Physics \& Astronomy, Princeton University, Princeton, NJ, 08544\\
$^{13}$Department of Physics \& Astronomy, Lehman College, City University of New York, 250 Bedford Park Blvd West, New York, NY 10468, USA\\
$^{14}$Institute of Physics, E\"otv\"os University, P\'azm\'any P.s., Budapest, 117, Hungary \\ 
$^{15}$Stanford Institute for Theoretical Physics, Stanford University, Stanford CA 94306, USA}

\begin{document}

\date{Accepted. Received; in original form}

\pagerange{\pageref{firstpage}--\pageref{lastpage}} \pubyear{2008}

\maketitle

\label{firstpage}

\begin{abstract}
We assess the contribution of dynamical hardening by direct three-body scattering interactions to the rate of stellar-mass black hole binary (BHB) mergers in galactic nuclei.  We derive an analytic model for the single-binary encounter rate in a nucleus with spherical and disk components hosting a super-massive black hole (SMBH).  We determine the total number of encounters $N_{\rm GW}$ needed to harden a BHB to the point that inspiral due to gravitational wave emission occurs before the next three-body scattering event.  This is done independently for both the spherical and disk components.  Using a Monte Carlo approach, we refine our calculations for $N_{\rm GW}$ to include gravitational wave emission between scattering events.  For astrophysically plausible models we find that typically  $N_{\rm GW} \lesssim$ 10.

We find two separate regimes for the efficient dynamical hardening of BHBs: (1) spherical star clusters with high central densities, low velocity dispersions and no significant Keplerian component; and (2) migration traps in disks around SMBHs lacking any significant spherical stellar component in the vicinity of the migration trap, which is expected due to effective orbital inclination reduction of any spherical population by the disk.  We also find a weak correlation between the ratio of the second-order velocity moment to velocity dispersion in galactic nuclei and the rate of BHB mergers, where this ratio is a proxy for the ratio between the rotation- and dispersion-supported components.  Because disks enforce planar interactions that are efficient in hardening BHBs, particularly in migration traps, they have high merger rates that can contribute significantly to the rate of BHB mergers detected by the advanced Laser Interferometer Gravitational-Wave Observatory. 

\end{abstract}


\section{Introduction}
Galactic nuclei have stellar densities that can reach $\sim 10^{7}$ M$_{\odot}$ pc$^{-3}$ \citep{Graham09}, inside some of the deepest gravitational potentials in the Universe. The resulting potential wells are multi-component, with significant contributions from a central supermassive black hole (SMBH), as well as the masses of stars, stellar remnants and gas. Stellar-mass black hole binaries (BHB) are believed to live in galactic nuclei due to a combination of stellar evolution and dynamical friction \citep{Morris93,Miralda00,antonini16}. If a central SMBH is present, compact remnants could even form a tightly packed cusp around it \citep[e.g.][]{Bartko09}. Plausible scenarios exist for the origins of BHBs in galactic nuclei, but how many to expect is largely unknown.  Some of these BHBs will harden to merger via dynamical encounters with other objects in the nucleus, emitting gravitational waves (GW) and yielding events detectable by the advanced Laser Interferometer Gravitational-wave Observatory (aLIGO) \citep[e.g.][]{konstantinidis13,leigh14,antonini16,leigh16a}. However, if a dense gas disk surrounds the SMBH, then in principle a fraction of the BHBs in the galactic nucleus can end up in the gas disk and can merge on short timescales. Thus, active galactic nucleus (AGN) disks could represent sites of efficient BHB mergers detectable with aLIGO \citep{McK14,Bartos17,Stone17}, possibly localized at migration traps in the disk \citep{Bello16}. 

This scenario was recently considered by \citet{Stone17}.  The authors found that stellar-mass BHBs can be driven to merge in Toomre-unstable AGN disks via a combination of three-body scattering events with background disk stars, gaseous torques from a circumbinary mini-disk and, eventually, GW emission (once the binary orbital separation is sufficiently small).  The authors argue that this mechanism should be most effective in small galaxies with SMBH masses of order 10$^6$-10$^8$ M$_{\odot}$.  \citet{Bartos17} similarly show that such BHB can merge effectively, aided by significant gas accretion from the disk at well above the Eddington rate.

Direct observational constraints for the properties of stellar and gaseous disks in galactic nuclei are scarce.  Close to home, stellar disks have been observed directly near the SMBHs in both the Milky Way (MW) and M31.  Within the core of the MW nuclear star cluster (NSC), of order a hundred OB stars rotate clockwise on the sky in a strongly warped disk $\sim$ 0.04-0.6 pc from the SMBH \citep[e.g.][]{levin03,lu09,Bartko10,Baruteau11,yelda14,chen15}.  The average stellar age and mass are inferred to be $\sim$ 6 Myr and $\gtrsim$ 10 M$_{\odot}$, respectively.  There exists tentative evidence for a secondary counterclockwise disk made up of stars similar in their photometric, spectroscopic and kinematic properties \citep[e.g.][]{Paumard06,Bartko09}.  In M31, there are two distinct nuclear disks.  A massive old disk of stars dominates the inner nucleus \citep{tremaine95}, with a total mass of $\sim$ 3 $\times$ 10$^7$ M$_{\rm \odot}$ and a half-power radius of $\sim$ 1.8 pc \citep{bender05}.  Inside this old disk and immediately surrounding the central SMBH is a compact nuclear disk of blue stars with a total mass $\sim$ 4200 M$_{\odot}$ and half-power radius $\sim$ 0.2 pc \citep{bender05,lauer98,lauer12}.

As for the gaseous counterparts to these stellar disks, AGN disks may be messy, sharing the large-scale structures evident in high resolution observations of protoplanetary disks \citep[e.g.][]{vandermarel13,perez14,casassus13,flock15}.  For example, water maser discs have been resolved on sub-parsec scales, revealing significant clumpiness \citep[e.g.][]{koekemoer95}.  
Consequently, some properties of AGN disks could be stochastic on short timescales.  We note the potential importance of these effects, since most theoretical studies neglect them due to their resistance to analytic modeling.  However, the stability of jets from quasars and radio galaxies argues for long-lived, stable inner disks, which is where we will focus much of our attention.

In a recent paper \citep{McK17}, hereafter Paper I, we parameterized the rate of BHB mergers expected from AGN disks. In Paper I we did not consider in detail the dynamical problem of BHB in the disk encountering tertiary objects, both in the disk and on orbits intersecting the disk, which will impact our merger rate estimate. Here we study the general problem of the rate of dynamical hardening of BHBs in a two-component nuclear star cluster around a SMBH. We assume the nuclear stellar cluster population is divided into two parts: a disk-like component and a spherical, bulge-like component.  Dynamical interactions are assumed to occur due to gravitational scattering events between both components. Implicit in our model is the maintenance of the disk-like component through the assumed existence of an AGN gas disk; in the absence of such gas, the disk-like component could be disrupted by the spherical component.\footnote{The destruction of the disk depends on the ratio between the timescale for vector resonant relaxation to operate and the timescale for orbital nodal precession mediated by the disk quadrupole \citep{rauch96,kocsis15}.  For disruption to occur, the former timescale needs to be shorter than the latter.} We do not explicitly model the NSC-gas interactions in this work, instead leaving it to future work.

In Section~\ref{model}, we begin by calculating the orbital separation beyond which an interloping single star will likely disrupt a BHB.  Next we calculate the rate of encounters between BHBs and single objects in both the disk and spherical components. Then we derive an analytic expression for the mean number of scattering interactions $N_{\rm GW}$ required to harden a BHB until the mean encounter timescale exceeds the inspiral timescale from GW emission. Finally we compare our analytic estimate for $N_{\rm GW}$ to the results of Monte-Carlo simulations that account for inter-encounter hardening and circularization due to GW emission.  Section~\ref{discussion} discusses and summarizes our results.

\section{Model} \label{model}

In this section, we present our model for the dynamical hardening of a BHB in a two-component galactic nucleus hosting a central SMBH.  That is, we assume that the Keplerian stellar disk (with or without gas) component is embedded within a spherical dispersion-supported stellar distribution.

\subsection{The hard-soft boundary} \label{HS}

In a star cluster composed of both single and binary stars, the "hard-soft" boundary is defined as the orbital separation of a typical binary with orbital energy roughly equal to the typical kinetic energy of a single star in the cluster. Binaries with orbital separations larger than the hard-soft boundary ("soft" binaries) tend to be disrupted or ionized during direct encounters with single stars in the cluster. Clusters with higher velocity dispersions have smaller orbital separations at the hard-soft boundary.

Specifically, a binary consisting of masses $M_{\rm 1},M_{\rm 2}$ with semi-major axis $a_{\rm b}$ has orbital energy 
$|E_{\rm orb}|=GM_{\rm 1}M_{\rm 2}/2a_{\rm b}$
and will be disrupted in a direct encounter where the single star passes within the binary orbit when $|E_{\rm orb}|<E_{\ast}=(1/2)M_{\ast}\sigma^{2}$, where $E_{\ast}$ is the typical kinetic energy of a tertiary encounter with $M_{\ast}$ the typical tertiary mass and $\sigma$ is the cluster velocity dispersion. Binaries with $|E_{\rm orb}|<E_{\ast}$ are soft, and the condition $|E_{\rm orb}|=E_{\ast}$ is the hard-soft boundary for binaries in the cluster, which sets a size scale:
\begin{equation}
\label{eqn:ahs0}
a_{\rm HS} = \frac{GM_{\rm 1}M_{\rm 2}}{M_{\rm *}\sigma^2}
\end{equation}

The hard-soft boundary for BHB will be different inside and outside the sphere of influence of the SMBH, characterized by the influence radius
\begin{equation}
r_{\rm inf}=\frac{GM_{\rm SMBH}}{\sigma^{2}} \approx 0.4\mbox{ pc} \left(\frac{M_{\rm SMBH}}{10^{7}\mbox{ M}_{\odot}}\right) \left(\frac{\sigma}{100 \mbox{ km s$^{-1}$}}\right)^{-2}
\label{eqn:rinf}
\end{equation}
where $M_{\rm SMBH}$ is the SMBH mass. For binaries located at radius $R_{b}>r_{\rm inf}$ from the SMBH, the hard-soft boundary can be parameterized as:
\begin{equation}
a_{\rm HS} \approx 18\mbox{ AU} \left( \frac{M_{\rm 1}}{10\mbox{ M}_{\odot}}\right) \left( \frac{M_{\rm 2}}{10\mbox{ M}_{\odot}}\right) \left( \frac{M_{\ast}}{0.5\mbox{ M}_{\odot}}\right)^{-1} \left( \frac{\sigma}{100{\mbox{ km s$^{-1}$}}}\right)^{-2}.
\label{eq:a_HS_NSC}
\end{equation}
BHB located at radius $R_{\rm b}>r_{\rm inf}$ with $a_{\rm b}<a_{\rm HS}$ will tend to harden with each successive tertiary encounter. 

For binaries located at radius $R_{\rm b}<r_{\rm inf}$, consider first those binaries in an AGN disk. The relative velocity $v_{\rm rel}$ at impact between single stars and binaries on co-planar orbits within the disk is
$v_{\rm rel}=v_{\rm orb}(R_{\rm b})-v_{\rm orb}(R_{\rm b}+a_{\rm HS,disk})$ for encounters with aligned orbital angular momentum, where: 
\begin{equation}
v_{\rm orb}=\left(\frac{GM_{\rm SMBH}}{R_{\rm b}}\right)^{1/2}
\end{equation}
is the Keplerian orbital velocity in the disk. Note that this equation for $v_{\rm rel}$ corresponds simply to the shear from one side of the binary to the other.  This assumes circular orbits, and hence no radial motion.\footnote{The assumption of circular orbits may not be realistical.  If the orbits of colliding single and binary stars are very eccentric, then this can increase the relative velocity at impact.  Hence, the relative velocities for circular orbits should be regarded as a minimum.}  Setting $(1/2) M_{\ast} v_{\rm rel}^{2} = GM_{\rm 1}M_{\rm 2}/2a_{\rm HS,disk}$ \citep{heggie75} and using a Taylor series expansion we find:
\begin{eqnarray}
a_{\rm HS,disk} &= & 2^{2/3}R_{\rm b} \left( \frac{M_{\rm 1}M_{\rm 2}}{M_{\ast}M_{\rm SMBH}}\right)^{1/3} \\
   & \simeq&(41 \mbox{ AU}) \left( \frac{R_{\rm b}}{0.01\mbox{ pc}}\right) \left( \frac{M_{\rm 1}}{10\mbox{ M}_{\odot}}\right) \left( \frac{M_{\rm 2}}{10\mbox{ M}_{\odot}}\right) \\
  &  &  \left( \frac{M_{\ast}}{0.5\mbox{ M}_{\odot}}\right)^{-1} \left( \frac{M_{\rm SMBH}}{10^8\mbox{ M}_{\odot}}\right)^{-1} 
\end{eqnarray}
which we can 
   also write 
as:
\begin{equation}
a_{\rm HS,disk}=12^{1/3} R_{\rm H} \left( \frac{\mu_{\rm b}}{M_{\ast}}\right)^{1/3}
\label{eq:a_HS_disk}
\end{equation}
where $R_{\rm H}=R_{\rm b}(q/3)^{1/3}$ is the Hill radius of the binary with $q=M_{\rm b}/M_{\rm SMBH}$ and $\mu_{\rm b}=M_{\rm 1}M_{\rm 2}/M_{\rm b}$ is the binary reduced mass.  Equation~(\ref{eq:a_HS_disk}) implies $a_{\rm HS,disk} \sim 12^{1/3} R_{\rm H}$ generally for $\mu_{\rm b}/M_{\ast} \sim 1$. 

For an encounter between a BHB and a tertiary mass (i.e., single star) from the spherical 
component orbiting at inclination angle $i$ relative to the orbit of the binary within the gas disk, the relative velocity at impact is: 
\begin{equation}
\label{eqn:vrel3}
v_{\rm rel}=\left( v^{2}_{\rm rel,\perp} + v^{2}_{\rm rel,\parallel}\right)^{1/2}
\end{equation}
where 
\begin{equation}
\label{eqn:vrel2}
v_{\rm rel, \perp}= v_{\rm orb}(R_{\rm b}) \sin (i)
\end{equation}
and
\begin{equation}
\label{eqn:vrel1}
v_{\rm rel, \parallel}= v_{\rm orb}(R_{\rm b}) -v_{\rm orb}(R_{\rm b}+a_{\rm HS,disk})\cos (i)
\end{equation}
In the limit $i \rightarrow 0^{\circ}$ and $a_{\rm HS,disk}/R_{\rm b} \ll 1$ we recover Equation~(\ref{eq:a_HS_disk}). As $i$ increases, the relative encounter velocity increases allowing for greater risk of 
       ionization.\footnote{We note in passing that it is straightforward to show that the Coriolis force at the Hill radius of a BHB is expected to be comparable to the acceleration towards the BHB. We defer a more thorough exploration of the impact of these effects on our results, using numerical scattering simulations, to a future study.}  We will show in Section~\ref{diskcomp} that objects in the spherical component at small disk radii have their orbital inclinations reduced by losing energy in the vertical direction at every passage through the disk until they reach coplanar orbits.  This process, which we refer to subsequently as grinding down of the stellar orbit, occurs relatively quickly.  As a result we can assume there are few out-of-plane encounters and so Equation~(\ref{eq:a_HS_disk}) is the appropriate hard-soft boundary for the AGN disk.  

\subsection{The three-body scattering rate}

Next, we calculate the rate of interactions between the BHB and stars in both the surrounding velocity-dispersion-supported component of the nuclear star cluster (NSC) and the disk component.  We assume that the former component resembles the NSC of the Milky Way.  

\subsubsection{The velocity-dispersion-supported component}

We begin with the velocity dispersion-supported stellar component of the NSC.  For the rate of collisions between the BHB and stars in this component, we adopt Equation (9) in \citet{leigh16a} along with their Equation (10) for the velocity dispersion with $\sigma_{\rm 0} =$ 100 km s$^{-1}$.  That is, the collision rate is given by:
\begin{equation}
\label{eqn:gamma10}
\Gamma_{\rm NSC} \approx ({\rho}\sigma + {\Sigma}\Omega)\left( \frac{R^2}{M} \right)\left(1 + \left( \frac{v_{\rm e}}{\sigma} \right)^2 \right),
\end{equation}
where $\rho$ and $\Sigma$ are the stellar (volume) mass density of the velocity-dispersion-supported component and the stellar surface mass density of the disk component, respectively, and $v_{\rm e}$ is the escape velocity from the target object (here the 
BHB; we take its semi-major axis as the relevant size).  The second term in Equation~(\ref{eqn:gamma10}) smoothly transitions the collision rate into and out of the regime where gravitational focusing becomes important.  The local velocity dispersion is given by:
\begin{equation}
\label{eqn:sigloc}
\sigma(r)^2 = \sigma_{\rm 0}^2 + \frac{GM_{\rm SMBH}}{r}.
\end{equation}  
Equation~(\ref{eqn:sigloc}) serves to modify the velocity dispersion very close to the central SMBH, where the stellar orbits begin transitioning into the Keplerian regime.

In calculating an order-of-magnitude estimate for the collision rate between the BHB and stars in the velocity-dispersion-supported component of the NSC, we ignore the disk component and set $\Sigma =$ 0 in Equation~(\ref{eqn:gamma10}).  We adopt the density profile given in Equations 1 and 4 of \citet{merritt10}, suitable to a velocity-dispersion-supported roughly isothermal nuclear environment.  The central density and half-mass radius (i.e., the distance from the cluster centre at which half the total stellar mass is enclosed) are taken to be 10$^6$ M$_{\odot}$ pc$^{-3}$ and 2.5 pc, respectively.  We adopt a mass-to-light ratio of 2, suitable to an old stellar population. 

\begin{figure}
\begin{center}
\includegraphics[width=\columnwidth]{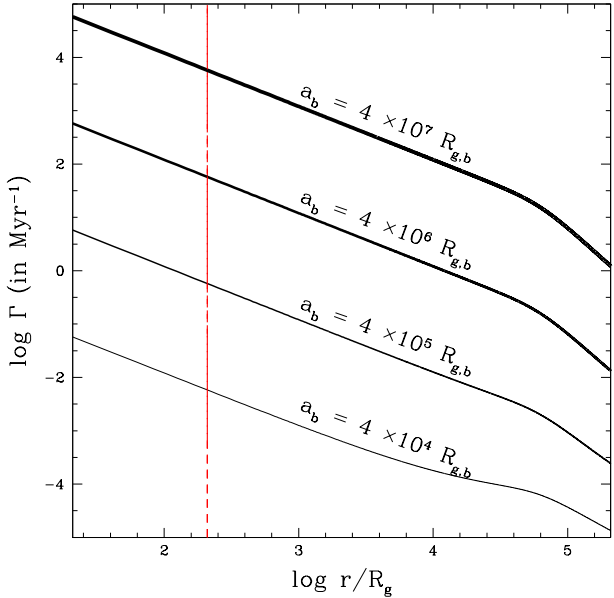}
\end{center}
\caption[The rate of direct encounters between the BHB and single stars in the spherical NSC as a function of distance from the SMBH]{The rate of collisions (in Myr$^{-1}$) between the BHB and single stars in the surrounding velocity-dispersion-supported component of the NSC, as a function of distance $r$ from the 
central SMBH (in units of the gravitational radius of the SMBH $R_{\rm g}$, with $M_{\rm SMBH} =$ 10$^8$ M$_{\odot}$).  We assume $m_{\rm BHB} =$ 25 M$_{\odot}$ for the BHB, but consider different orbital separations (as shown above each line in the figure).  The line thickness is directly proportional to the orbital separation of the BHB, and we consider orbital separations of $a_{\rm BHB} =$ 4 $\times$ 10$^4$, 4 $\times$ 10$^5$, 4 $\times$ 10$^6$, 4 $\times$ 10$^7$
      (0.2--20~AU for this binary mass), 
in units of the gravitational radius of the binary $R_{\rm g,b}$.  The dashed vertical red line shows the location of the migration trap 
based on a calculation by \citet{Bello16} using the disk model of \citet{thompson05}.
\label{fig:fig2}}
\end{figure}

Figure~\ref{fig:fig2} shows the rate of collisions between the BHB and other single stars in the velocity-dispersion-supported component of the NSC, for different assumptions about the BHB orbital separation,
     measured in units of the gravitational separation of the binary
     \begin{equation}
           R_{g,b} = \frac{2 G M_{BHB}}{c^2} = (4.9 \times 10^{-7} \mbox{ AU}) \left(\frac{M_{BHB}}{25 \mbox{ 
           M}_{\odot}}\right) .
      \end{equation}
We assume a BHB mass of 25 M$_{\odot}$ and SMBH mass of 10$^8$ M$_{\odot}$.  
    These collisions will, however, be disruptive for all but the hardest BHBs, which have the 
    lowest encounter rates.  

\subsubsection{The disk component} \label{diskcomp}

Binaries interacting with an isotropic distribution of stars 
   such as the velocity-dispersion-supported component 
are most likely to be dissociated by those interactions, due to the large relative velocities of the encounters, predominantly at high inclination angles. Thus, the high collision rates shown in Figure~\ref{fig:fig2} will not lead to a high BHB merger rate. 
    By contrast, most stars orbiting within $\lesssim$ 10$^3$$R_{\rm g}$ of the central SMBH to 
    will have their orbital inclinations reduced by the disk in less than a megayear 
    \citep{artymowicz93}, much less than the disk lifetime. As a result, BHBs in this region will
predominantly encounter stars in co-planar orbits.  In this case \citep{artymowicz93}, the relative encounter velocities can be very low (see Equations~(\ref{eqn:vrel3}), ~(\ref{eqn:vrel2}) and~(\ref{eqn:vrel1}) with $i \sim$ 0$^{\circ}$), 
      so
disruption is unlikely 
     \citet[see][for the protoplanetary case]{horn12}, and encounters
      can effectively harden BHBs orbiting there.


    Furthermore, as discussed in Paper I, BHBs can migrate to or form in one of the migration traps that \citet{Bello16} showed 
     occur
in AGN disks. 
     For example, 
\citet{Bello16} found a migration trap at a distance of $r_{\rm trap} =$ 200$R_{\rm g}$ from the central SMBH 
    using the disk model of \citet{thompson05}, where $R_{\rm g} =$ 2G$M_{\rm SMBH}$/c$^2$ is the gravitational radius of the SMBH. 

We wish to obtain a crude estimate of the rate of encounters between a BHB 
   in a trap
and single stars migrating within the AGN disk. 
   We
assume all stars from the spherical component have had their orbits sufficiently dissipated that they are co-planar with the disk (see Section~\ref{migration} for more details).  We use the Type I migration time-scale appropriate to a prograde orbit about the central SMBH \citep{Paarde10}:
\begin{equation}
\begin{aligned}
\label{eqn:timeI}
\tau_{\rm I} = 19 \mbox{ Myr} \left( \frac{\beta}{3} \right)^{-1} \left( \frac{r}{10^4 r_{\rm g}} \right)^{-1/2} \left( \frac{M_{\rm *}}{10\mbox{ M}_{\odot}} \right)^{-1} \\
\left( \frac{h/r}{0.02} \right)^2 \left( \frac{\Sigma}{10^5\mbox{ M$_{\odot}$ pc$^{-2}$}} \right)^{-1} \left( \frac{M_{\rm SMBH}}{10^8\mbox{ M}_{\odot}} \right)^{3/2},
\end{aligned}
\end{equation}
where $\beta$ is a numerical factor of order 3, $h$ is the disk scale height and $r$ is the distance from the central SMBH.\footnote{We caution that inside $\lesssim$ 0.01 pc (i.e., $\lesssim$ 10$^3$ $R_{\rm g}$),
the gap-opening criteria can be satisfied in the disk \citep{Lin93,Baruteau11,Stone17,Bartos17} and the Type II migration time could become more relevant.}  We adopt $h/r =$ 0.02 for the disk aspect ratio.  This is a reasonable choice for a "typical" disk, given that the aspect ratio $h/r$ of a disk can vary considerably among disk models.  For example, \citet{thompson05} estimate $h/R \sim$ 10$^{-3}$--0.1, whereas \citet{sirko03} estimate $h/R \sim$ 0.01--0.1 in their models.  Equation (\ref{eqn:timeI}) shows that for a very thin disk, the inspiral rate is two orders of magnitude higher than in our fiducial model.  For a very thick advection dominated accretion flow (ADAF), the inspiral rate is two orders of magnitude lower relative to our fiducial case.  Thus, uncertainties in the disk aspect ratio translate 
into considerable uncertainty in our final estimate for the encounter rate in the disk, which could span up to four orders of magnitude due to this source of uncertainty alone.  We will return to this important issue in Section~\ref{discussion}.  

Assuming that the rate of migration is constant, the mean rate of inward migration can then be estimated as:
\begin{equation}
\label{eqn:drdt}
\frac{dr}{dt}(r) \sim \frac{r_{\rm out}}{\tau_{\rm I}(r = r_{\rm out})}.
\end{equation}
Now, assuming $N_{\rm *} =$ 10$^3$ stars in the disk (i.e., 10$^{-3} M_{\rm disk}$ for a total gas mass $M_{\rm gas} =10^6$ M$_{\odot}$, 
the mean inter-star spacing in the disk is ${\Delta}r \sim r_{\rm out}/N_{\rm *} \sim 10 R_{\rm g} \sim 10$ AU, where $r_{\rm out} = 10^4 R_{\rm g}$ is the outer limiting radius of the disk \citep{McK17} and $R_{\rm g}$ is the gravitational radius of the central SMBH.  Thus, the time between scattering events with disk stars is roughly:
\begin{equation}
\label{eqn:timedisk}
\tau_{\rm disk}(r) \sim \frac{{\Delta}r}{dr/dt} \sim \frac{\tau_{\rm I}(r_{\rm out})}{N_{\rm *}}
\end{equation}
The corresponding encounter rate is:
\begin{equation}
\label{eqn:gamma_disk}
\Gamma_{\rm disk}(r) = \frac{1}{\tau_{\rm disk}}
\end{equation}
In scaled form, Equation~(\ref{eqn:gamma_disk}) becomes:
\begin{equation}
\begin{aligned}
\label{eqn:gamma_disk2}
\Gamma_{\rm disk}(r) = 53 \mbox{ Myr}^{-1} \left( \frac{\beta}{3} \right) \left( \frac{r}{10^4 r_{\rm g}} \right)^{1/2} \left( \frac{M_{\rm *}}{10\mbox{ M}_{\odot}} \right) \\
\left( \frac{h/r}{0.02} \right)^{-2} \left( \frac{\Sigma}{10^5\mbox{ M$_{\odot}$ pc$^{-2}$}} \right) \left( \frac{M_{\rm SMBH}}{10^8\mbox{ M}_{\odot}} \right)^{-3/2} \left( \frac{N_{\rm *}}{10^3} \right)
\end{aligned}
\end{equation}
We can use Equation (\ref{eqn:timeI}) to estimate the scattering encounter timescale $\tau_{\rm disk}$ for stars with $M_{\ast} = 1 \mbox{ M}_{\odot}$ at $r_{\rm trap}$, migrating in from $r = r_{\rm out}$, and assuming $\Sigma \sim 10^7$ M$_{\odot}$ pc$^{-2}$ 
with the other parameters given to be $\tau_{disk} = 1.9$~kyr, so $\Gamma_{\rm disk}(r_{\rm trap}) = 5.3 \times 10^2 \mbox{ Myr}^{-1}$.

We emphasize that our estimate for the rate of encounters between the BHB and stars in the AGN disk is very rough, due in large part to the lack of strong observational constraints for the properties of AGN disks, but also the analytic limitations of our model, 
not least of which is approximating the radially variable rate of accretion as a constant.\footnote{We assume constant or average values for several parameters characteristic of AGN disks in Equation~\ref{eqn:timeI}.  Consequently, our calculated estimate for this timescale is approximate, and falls below the maximum value calculated by \citet{thompson05}.  Regardless, this discrepancy does not affect our results our for BHB hardening in AGN disks, since our estimate for this rate could decrease by more than three orders of magnitude without affecting our conclusions.  This remains well within our uncertainty in our estimate for the migration timescale.}  
The take-away message though is that this timescale can be very short, relative to the timescale for encounters with single stars in the absence of the dissipative forces supplied by a gaseous disk.

\subsection{Scattering number for binary merger} \label{num}

In this section, we derive an analytic expression for the typical number $N_{\rm GW}$ of direct three-body scattering interactions needed to harden a BHB to the point that the timescale for the next encounter to occur exceeds the timescale for the BHB to merge via GW emission.  This number is calculated in the absence of GW emission, and represents the energy loss due to gravitational scattering alone.  Hence, including GW emission in the rate of binary hardening will only decrease this number, relative to the analytic expression derived here.  In the limit that GW emission leads to rapid inspiral in a time less than the time for even a single scattering interaction to occur, then $N_{\rm GW} \rightarrow$ 0.

Following \citet{valtonen06}, we introduce the distribution of single star escaper velocities for chaotic three-body interactions with total encounter energy $E_{\rm 0}$.  From Equation (7.19) in \citet{valtonen06}, we have for the distribution of single star escaper velocities produced during single-binary interactions:
\begin{equation}
\label{eqn:fvs}
f(v_{\rm s})dv_{\rm s} = \frac{\left( (n-1)|E_{\rm 0}|^{n-1}m_{\rm s}M/m_{\rm B} \right)v_{\rm s}dv_{\rm s}}{\left( |E_{\rm 0}| + \frac{1}{2}(m_{\rm s}M/m_{\rm B})v_{\rm s}^2 \right)^n},
\end{equation}
where $m_{\rm s}$ is the mass of the escaper, $m_{\rm B}$ is the binary mass and $M =$ $m_{\rm s} +$ $m_{\rm B}$ is the total system mass.  This equation gives the normalized probability of obtaining a given escaper velocity $v_{\rm s}$ for a given initial encounter energy $E_{\rm 0}$.  The index $n$ accounts for the angular momentum dependence, and for the three-body problem it is given by \citep{valtonen06}:
\begin{equation}
\label{eqn:n}
n - 3 = 18L^2,
\end{equation}
where we normalize the angular momentum L$_{\rm 0}$ according to $L =$ $L_{\rm 0}$/$L_{\rm max}$ such that 0 $\le L$ $\le$ 1 and:
\begin{equation}
\label{eqn:Lmax}
L_{\rm max} = 2.5G\left( \frac{m_{\rm 0}^5}{|E_{\rm 0}|} \right)^{1/2},
\end{equation}
where
\begin{equation}
\label{eqn:m0}
m_{\rm 0} = \left( \frac{m_{\rm a}m_{\rm b} + m_{\rm a}m_{\rm s} + m_{\rm b}m_{\rm s}}{3} \right)^{1/2},
\end{equation}
and $m_{\rm B} =$ $m_{\rm a} +$ $m_{\rm b}$.

We can differentiate Equation~(\ref{eqn:fvs}) with respect to $v_{\rm s}$, and set the result equal to zero to solve for the mode of the escape velocity distribution:
\begin{equation}
\label{eqn:vspeak}
v_{\rm s,mode} = \alpha \left( \frac{M-m_{\rm s}}{m_{\rm s}M}|E_{\rm 0}| \right)^{1/2},
\end{equation}
where
\begin{equation}
\label{eqn:n2}
\alpha = \left(n - \frac{1}{2} \right)^{-1/2}
\end{equation}
We use the mode of the velocity distribution, instead of the mean, because this provides a simple analytic equation for the number of hardening interactions.  For the mean, we must instead integrate numerically to compute its value.  We have performed this exercise, and find that our results do not change significantly from using the mode.

Using conservation of energy, we can write the final binary orbital energy of the BHB after an interaction with initial energy $E$ as
\begin{equation}
\label{eqn:Ebf}
E_{\rm B,f} = E - (K_{\rm s,f} + K_{\rm B,f}),
\end{equation}
where $K_{\rm s,f}$ and $K_{\rm B,f}$ are the kinetic energies of the escaping single star and binary, respectively, with respect to the 
system centre of mass.  We set $K_{\rm s,f} = \frac{1}{2}m_{\rm s}v_{\rm s,mode}^2$ and $K_{\rm B,f} = \frac{1}{2}m_{\rm B}v_{\rm B}^2$.  The final binary velocity is found using conservation of linear momentum, or by setting $v_{\rm B} =$ ($m_{\rm s}$/$m_{\rm B}$)$v_{\rm s,mode}$.  

At the start of a series of encounters, the total encounter energy $E_{\rm i}$ 
of the first single-binary scattering event is
\begin{equation}
\label{eqn:E0}
E_{\rm i} = K_{\rm s,i} + K_{\rm B,i} + E_{\rm B,i} = \frac{1}{2}m_{\rm s}v_{\rm s,i}^2 + \frac{1}{2}m_{\rm B}v_{\rm B,i}^2 + E_{\rm B,i},
\end{equation}
where $v_{\rm s,i}$ and $v_{\rm B,i}$ are, respectively, the initial velocities of the single star and the binary relative to the system centre of mass.
Now, the energy exchanged per subsequent interaction is:
\begin{equation}
\label{eqn:dEdN}
\frac{dE}{dN} = E - (K_{\rm s,f} + K_{\rm B,f}),
\end{equation}
where $E$ is the total encounter energy of each of the subsequent three-body scattering events in turn.  With each interaction, we perform an operation to subtract off the total positive energy the interacting three-body system (including the incoming single star) loses, $K_{\rm s,f} + K_{\rm B,f}$. 

Our goal is to compute how many interactions must occur to drain the binary energy $E_B$ to a level such that the system energy $E$ has dropped to its desired final total encounter energy, 
\begin{equation}
\label{eqn:Ef}
E_{\rm f} = E_{\rm GW} + K_{\rm s,f} + K_{\rm B,f}.
\end{equation}  
where $E_{\rm B,f} =$ $E_{\rm GW}$ corresponds to the critical binary orbital energy at which the timescale for a merger due to GW emission becomes less than the timescale for a subsequent scattering interaction to occur.  

The required total number of interactions needed to harden a BHB to a final orbital energy $E_{\rm B,f}$ is:
\begin{equation}
\label{eqn:numhard}
N = \int_0^N dN = \int_{E_{\rm i}}^{E_{\rm f}} \left( \frac{dE}{dN} \right)^{-1}dE = \int_{E_{\rm i}}^{E_{\rm f}} \frac{dE}{E - (K_{\rm s,f} + K_{\rm B,f})},
\end{equation}
This can be re-written as:
\begin{equation}
\label{eqn:numhard2}
N = \int_{E_{\rm i}}^{E_{\rm f}} \frac{dE}{E(1 - \frac{1}{2}m_{\rm s}\alpha^2((M-m_{\rm s})/(m_{\rm s}M)(1+m_{\rm s}/m_{\rm B}))},
\end{equation}
where we have used conservation of linear momentum, $v_{\rm B} =$ ($m_{\rm s}$/$m_{\rm B}$)$v_{\rm s}$.  Integrating Equation~(\ref{eqn:numhard2}) gives:
\begin{equation}
\label{eqn:numhard3}
N = \left( 1 - \frac{1}{2}m_{\rm s}\alpha^2\left( \frac{M-m_{\rm s}}{m_{\rm s}M} \right)\left(1 + \frac{m_{\rm s}}{m_{\rm B}} \right) \right)^{-1}{\ln}\left( \frac{E_{\rm f}}{E_{\rm i}} \right)
\end{equation}
 
For interactions with stars in the velocity-dispersion-supported component, the interactions are distributed isotropically and the corresponding eccentricity distribution is thermal:
\begin{equation}
\label{eqn:ecc1}
f(e)de = 2ede
\end{equation}

What about encounters with disk stars, which are restricted to occur in the orbital plane of the binary?  In the planar case, the distribution of orbital energies $E_{\rm B}$ and eccentricities $e$ after a single encounter are, respectively \citep{valtonen06}:
\begin{equation}
\label{eqn:Eb2}
f(|E_{\rm B}|)d|E_{\rm B}| = 2|E_{\rm 0}|^2|E_{\rm B}|^{-3}d|E_{\rm B}|,
\end{equation}
and
\begin{equation}
\label{eqn:ecc2}
f(e)de = e(1-e^2)^{-1/2}de
\end{equation}
As described in more detail in Section 7.2 of \citet{valtonen06}, Equation~(\ref{eqn:ecc2}) can be derived using a statistical mechanics approach first pioneered by \citet{monaghan76a} and \citet{monaghan76b}, and has been verified using numerical orbital calculations \citep[e.g.][]{saslaw74}.  Hence, $n =$ 3 and $\alpha = \sqrt{2/5}$ from Equations~(\ref{eqn:n}) and~(\ref{eqn:n2}). 

The solid lines in Figure~\ref{fig:fig3} show the range in the typical number of encounters $N_{\rm GW}$ needed to harden a BHB to the point of rapid inspiral as a function of the initial orbital separation.  For this exercise, we assume $M_{\rm SMBH} =$ 10$^6$ M$_{\odot}$, and place the BHB at a distance from the central SMBH equal to $r_{\rm inf}$ (Equation~(\ref{eqn:rinf})).  Recall that encounters with stars in the surrounding velocity-dispersion-supported stellar component should have random angular momenta (and hence a range of values for $n$ in Equation~(\ref{eqn:n}); shown by the black and red lines), whereas encounters with stars in the disk component should have low angular momentum (and hence $n =$ 3 in Equation~(\ref{eqn:n}); shown in blue).  From Figure~\ref{fig:fig3}, this suggests that encounters with stars from the spherical component of the NSC are less efficient at hardening the BHB than are encounters with disk stars.  This agrees with the results of \citet{Stone17}.  \textit{To conclude, the range in $N_{\rm GW}$ is narrow; of order one to ten encounters are needed almost independent of the total encounter angular momentum.} 

\subsection{Eccentricity and angular momentum evolution} \label{ecc}

Given that the timescale for inspiral due to GW emission scales steeply with the orbital eccentricity, understanding the evolution of the binary eccentricity as it hardens due to dynamical interactions can be crucial.  We stress that our analytic formulae for the critical number of interactions derived in the preceding sections do not include hardening due to GW emission that can occur in between encounters.  Consequently, they correspond to upper limits for the critical number of hardening interactions.

To account for this and test the validity of our analytic formulae, we run a series of Monte Carlo simulations for the evolution of BHBs hardening in a NSC with both velocity-dispersion-supported and disk components.  We describe in detail our Monte Carlo simulations for these two environments separately in the subsequent sections.  Interactions between BHBs and cluster stars can have any angular momentum, whereas encounters with disk stars are assumed to be confined to the orbital plane of the binary.  For this exercise, we consider two locations for our BHBs, namely the influence radius of the NSC (Equation~(\ref{eqn:rinf})), and a migration trap within the AGN disk.

\subsubsection{The influence radius} \label{rinf}

We assume in this section an idealized stellar disk that extends out to the influence radius.  This facilitates a comparison of hardening due to interactions with disk stars relative to cluster stars, and hence planar versus isotropic scatterings.  We consider the disk and spherical components separately.  Again, this is an over-simplification adopted for illustrative purposes.  However, as we will show, a comparison of these two extremes can be used to quantify the effects of anisotropic velocity distributions on the dynamical hardening of BHBs.

\begin{figure}
\begin{center}
\includegraphics[width=\columnwidth]{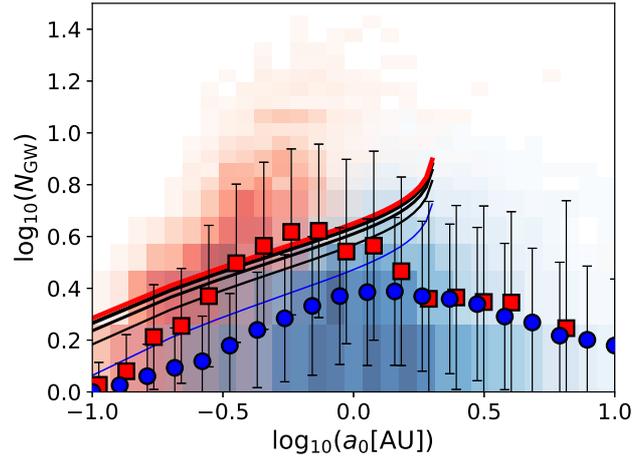}
\end{center}
\caption[The number of hardening interactions $N_{\rm GW}$ needed to harden a BHB to the point of rapid inspiral for a central cluster density of 10$^6$ M$_{\odot}$ pc$^{-3}$ without perturbations]{The solid lines show the analytic estimate of Equation~(\ref{eqn:numhard3}) for the number of hardening encounters $N_{\rm GW}$ needed to harden a BHB from a given initial orbital separation $a_{\rm 0}$ to a final orbital separation $a_{\rm GW}$ at which the inspiral time due to GW emission is shorter than the timescale for the next encounter.  We consider different values for the total encounter angular momentum by varying the parameter $n$ in Equation~(\ref{eqn:n}) from $n =$ 3 to $n =$ 7.  The line thickness is proportional to $n$.  The blue line shows $n =$ 3, appropriate to disk interactions.  The red line shows $n =$ 7, appropriate to interactions with stars in the surrounding velocity-dispersion-supported spherical stellar distribution (see text).  The solid lines end at the hard-soft boundary beyond which $E > 0$ for the initial relative velocities set by the chosen velocity dispersion.  Results of our Monte Carlo simulations neglecting perturbing encounters (see text) are shown for stars in the spherical (red shaded region) and disk (blue shaded region) components of the NSC.  The colour intensity is proportional to the number of simulations run in that bin of orbital separation with the given $N_{\rm GW}$.  The solid circles and squares show for planar and isotropic scattering, respectively, the mean number of hardening encounters for BHBs that merge for each bin in the initial binary orbital separation, using bin sizes of 1 AU.  
\label{fig:fig3}}
\end{figure}

Thus, we consider two different sets of simulations.  The first set of simulations is applicable to encounters with stars in the velocity-dispersion-supported (i.e., isotropic) component.  We sample the parameter $n$ in Equation~(\ref{eqn:n}) uniformly between $n =$ 3 and $n =$ 14.5 for every encounter, using the eccentricities from Equation~(\ref{eqn:ecc1}).  Note that in Figure~\ref{fig:fig3}, we take $n = 7$ as being representative of this range (in part to avoid over-crowding since the lines converge for larger $n$) .  The second set of simulations is applicable to encounters with stars in a disk (i.e., planar); we fix $n =$ 3 in Equation~(\ref{eqn:n}) and sample the eccentricities from Equation~(\ref{eqn:ecc2}).  

For each realization, we sample the initial binary orbital separation uniformly in the range 0--10 AU.  All binaries have component masses of 10 M$_{\odot}$ and 15 M$_{\odot}$, and all interloping single stars have masses of 1 M$_{\odot}$.  That is, we take $m_{\rm s} =$ 1 M$_{\odot}$, $m_{\rm a} =$ 10 M$_{\odot}$ and $m_{\rm b} =$ 15 M$_{\odot}$ in Equation~(\ref{eqn:numhard3}).  We assume a relative velocity at infinity of 100 km s$^{-1}$ for all encounters, and we set $a_{\rm GW} =$ 0.01 AU since this is less than the critical separation at which the inspiral time drops below the encounter time (our results are insensitive to the exact choice for $a_{\rm GW}$).  Finally, to calculate the rate of single-binary encounters, we adopt the density profile given by Equations 1 and 4 in \citet{merritt10} for the velocity-dispersion-supported component with a central density of 10$^6$ M$_{\odot}$ pc$^{-3}$, a half-mass radius of $r_{\rm h} =$ 2.5 pc \citep{merritt13} and a mass-to-light ratio of 2.  For simplicity, we adopt this same encounter rate for interactions with disk stars, to better isolate the effects of individual scattering events in the isotropic and planar regimes on the overall hardening rate.

Given these initial parameters, we calculate the time for another single star to encounter the BHB, for both the velocity-dispersion-supported and disk components.\footnote{We use the analytic single-binary (1+2) encounter time given in \citet{leigh11} for both the velocity-dispersion-supported and disk components, even though this formula is only appropriate to a spherical stellar distribution.  In this way, our Monte Carlo simulations are designed to quantify the effects of only the total angular momentum characteristic of the 1+2 interactions in deciding the final inspiral time of the BHB.}  We then evolve the binary orbital separation and eccentricity forward in time using the formulae in \citet{Peters64}, which account for the effects of GW emission, for a total time equal to the calculated single-binary encounter time $\tau_{\rm 1+2}$.  At this point, a new 1+2 encounter time is calculated, and the process is repeated until the BHB orbital separation and eccentricity yield an inspiral time $\tau_{\rm GW}$ (also given in \citet{Peters64}) that is shorter than the tertiary encounter time (i.e., $a_{\rm GW} \lesssim$ 0.01 AU).  Thus, for each value of the initial BHB orbital separation, we end up with a total number of encounters $N_{\rm GW}$ needed to satisfy $\tau_{\rm GW} < \tau_{\rm 1+2}$.   All encounters in these realizations of our Monte Carlo approach are assumed to be direct.  To explore the effects of perturbing encounters, we include additional Monte Carlo simulations in which we assume that two perturbing encounters occur for every direct encounter.  That is, guided by Figure 4 in \citet{leigh16c} which illustrates the significance of perturbing encounters as a function of the impact parameter using numerical scattering simulations, we sample a new binary eccentricity twice as frequently as we sample a new binary orbital energy.  This is because the change in eccentricity remains high relative to the change in orbital separation due to perturbing encounters for larger impact parameters, by up to a factor of about two \citep{leigh16c}.  In total, for both the velocity-dispersion-supported and disk components, we perform 10$^5$ realizations of our Monte Carlo simulations.

The results of the above analysis are shown by the red squares and blue circles in Figure~\ref{fig:fig3}, each of which corresponds to the binned mean of our Monte Carlo simulations at the indicated initial orbital separation, for which we adopt bin sizes of 1 AU.  The red and blue shaded regions show the distributions for all 10$^5$ realizations of our Monte Carlo simulations, with the colour intensity being proportional to the number of simulations at the indicated initial BHB orbital separation and $N_{\rm GW}$.  The error bars are calculated assuming Poisson statistics.  Note that, if the sampled total encounter energy is positive, we assume that the BHB is dissociated and never undergoes a merger, and only include BHBs that merge in calculating the mean numbers of encounters for each bin in the initial separation.  As shown in Table~\ref{table:stats}, dissociation occurs more commonly for interactions with isotropic cluster stars, due to the lower peak single star escape velocities.  This leaves the BHB more vulnerable to dissociation during a second 1+2 interaction.  

We can draw two conclusions from Figure~\ref{fig:fig3}:  (1) The analytic estimate for the number of hardening interactions $N_{\rm GW}$ given in Equation~(\ref{eqn:numhard3}) over-estimates the true number by a factor of a few when inter-encounter hardening due to GW emission is taken into account; and (2) interactions with stars in the velocity-dispersion-supported or spherical component are more likely to dissociate BHBs relative to interactions with disk stars.  This is because the lower angular momenta characteristic of interactions with disk stars yield higher typical escaper velocities and higher (positive) escaper kinetic energies.  This increases the probability that the interaction will significantly harden the binary during the first scattering event, thus decreasing the probability of dissociation during a second encounter.  We note that there is some evidence that Equation~\ref{eqn:fvs} over-predicts the peak of the escaper velocity distribution at high virial ratios \citep{leigh16b}.  This would contribute to a slight under-estimate of the total number of hardening interactions $N_{\rm GW}$, however this affects our results by at most a factor of order unity.

\begin{figure}
\begin{center}
\includegraphics[width=\columnwidth]{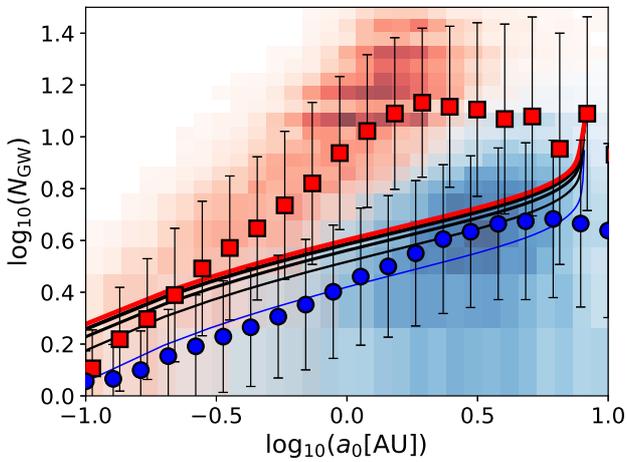}
\end{center}
\caption[The number of hardening interactions $N_{\rm GW}$ needed to harden a BHB to the point of rapid inspiral for a cluster density and velocity dispersion of, respectively, 10$^7$ M$_{\odot}$ pc$^{-3}$ and 50 km s$^{-1}$, without perturbations]{Same as Figure~\ref{fig:fig3} but assuming a factor of ten higher stellar mass density of $\rho_{\rm 0} =$ 10$^7$ M$_{\odot}$ pc$^{-3}$, a factor of two lower stellar velocity dispersion of $\sigma =$ 50 km s$^{-1}$ and no perturbing encounters.  
\label{fig:fig4}}
\end{figure}

The first point is further illustrated in Figure~\ref{fig:fig4}, for which we assume a factor of ten higher central stellar mass density of $\rho =$ 10$^7$ M$_{\odot}$ pc$^{-3}$, and a factor of two lower velocity dispersion of $\sigma =$ 50 km s$^{-1}$.  The higher density translates into a shorter encounter time.  This increases the efficiency of dynamical hardening relative to the effects of gravitational wave emission, and causes the simulated data points to shift upward relative to our analytic predictions.  The lower velocity dispersion also translates into a larger orbital separation at the hard-soft boundary.  The second point is illustrated in Table~\ref{table:stats}, which shows that a larger fraction of interactions in the spherical or cluster component disrupt BHBs relative to interactions in the disk component.  The fraction of disrupted interactions should increase as BHBs migrate inward from the influence radius, at least until they are sufficiently far in that all stellar orbits have been ground down in to the disk and high inclination encounters (with a relative velocity of order the local Keplerian speed) become unlikely.

\begin{figure}
\begin{center}
\includegraphics[width=\columnwidth]{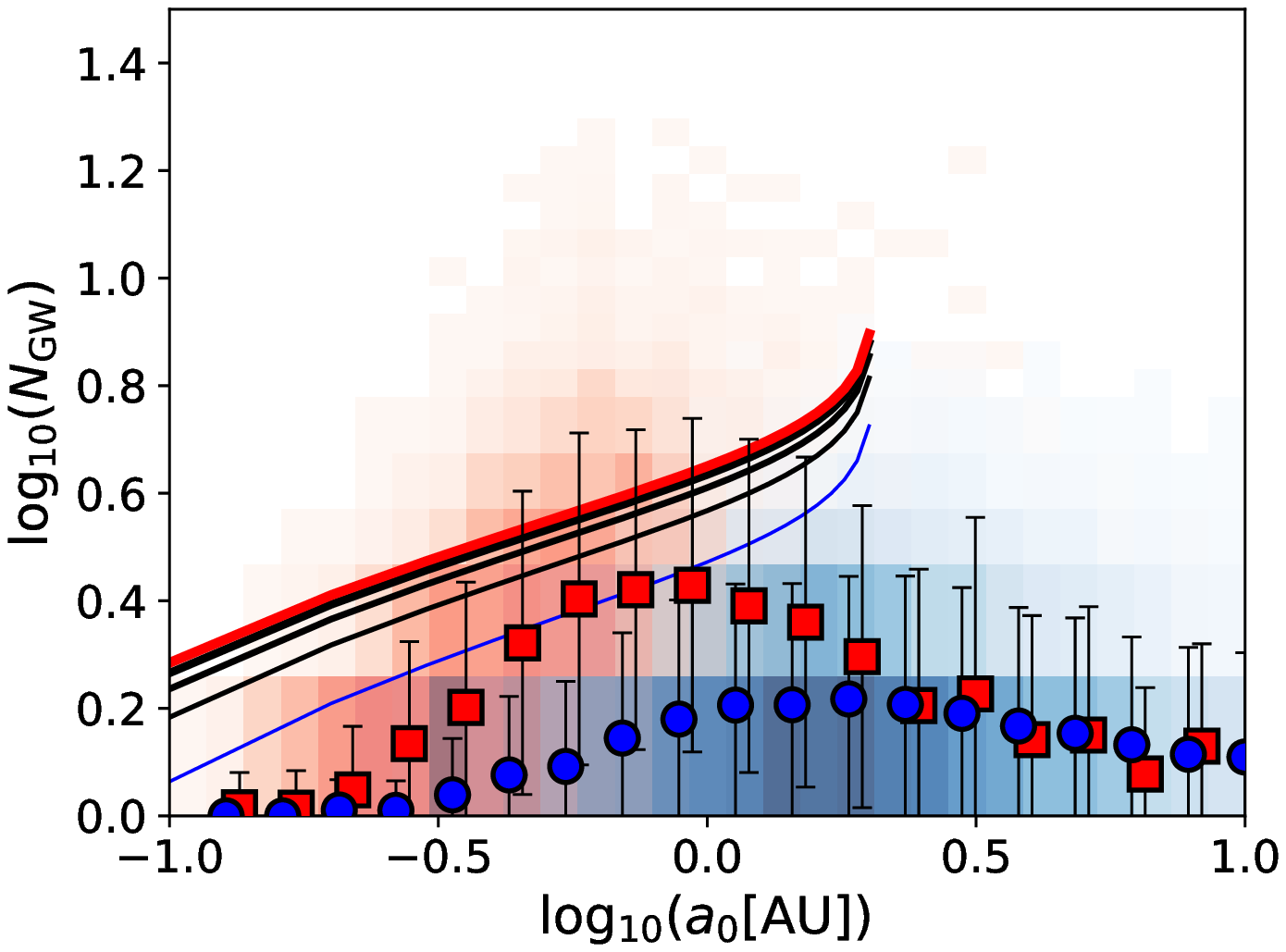}
\end{center}
\caption[The number of hardening interactions $N_{\rm GW}$ needed to harden a BHB to the point of rapid inspiral for a cluster density and velocity dispersion of, respectively, 10$^6$ M$_{\odot}$ pc$^{-3}$ and 100 km s$^{-1}$, with perturbations]{Same as Figure~\ref{fig:fig3} but including perturbing encounters.
\label{fig:fig5}}
\end{figure}

\begin{figure}
\begin{center}
\includegraphics[width=\columnwidth]{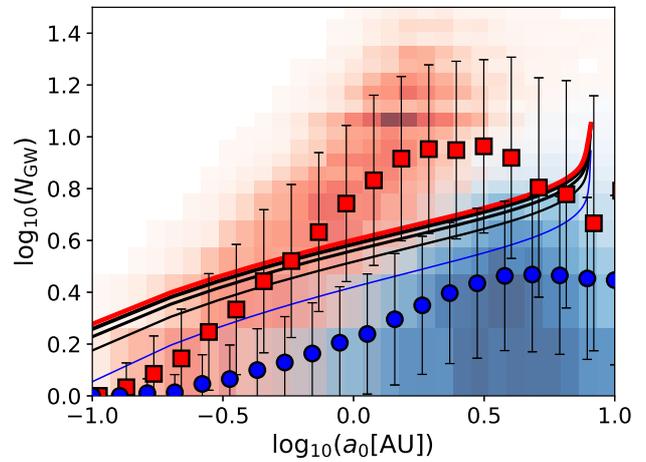}
\end{center}
\caption[The number of hardening interactions $N_{\rm GW}$ needed to harden a BHB to the point of rapid inspiral for a cluster density and velocity dispersion of, respectively, 10$^7$ M$_{\odot}$ pc$^{-3}$ and 50 km s$^{-1}$, with perturbations]{Same as Figure~\ref{fig:fig4} but including perturbing encounters.
\label{fig:fig6}}
\end{figure}

Finally, Figures~\ref{fig:fig5} and~\ref{fig:fig6} illustrate the effects of including perturbing encounters in our Monte Carlo simulations.  These serve to increase the binary orbital eccentricity, and hence to increase the efficiency of inter-encounter GW emission.  As a result $N_{\rm GW}$ drops further compared to the analytic estimate.


\begin{table*}
\caption{The numbers of BHBs that either merge or are disrupted in our Monte Carlo simulations.}
\begin{tabular}{|c|c|c|c|c|c|c|c|}
\hline
Density      &     Velocity Dispersion  &  \multicolumn{2}{|c|}{Perturbations}  &  \multicolumn{2}{|c|}{Cluster}    &   \multicolumn{2}{|c|}{Disk}   \\
  (M$_{\odot}$ pc$^{-3}$)   &   (km s$^{-1}$)   &   With    &   Without     &     Merged    &    Disrupted    &     Merged    &     Disrupted \\
    \hline
10$^5$     &     20   &      &   x   &   95219   &   4781   &   99989   &   11    \\
10$^5$     &     50   &      &   x   &   22450   &   77550   &   58450   &  41550    \\
10$^5$     &   100   &      &   x   &    7706   &    92294   &   22780   &  77220    \\
10$^5$     &   250   &      &   x   &    2838   &    97162   &   10160   &  89840    \\
10$^6$     &     20   &      &   x   &   94252   &    5748   &    99980  &        20    \\
10$^6$     &     50   &      &   x   &   20930   &   79070   &   56700  &   43300    \\
10$^6$     &   100   &      &   x   &    6332   &   93668   &   20200  &   79800    \\
10$^6$     &   100   &  x    &      &   8586   &   91414   &   30450   &  69550    \\
10$^6$     &   250   &      &   x   &    1828   &   98172   &    7908  &   92092    \\
10$^7$     &     20   &      &   x   &   93765   &    6235   &   99975  &       25    \\
10$^7$     &     50   &      &   x   &    20240   &   79760   &   54960  &   45040    \\
10$^7$     &     50   &  x    &      &   21360   &   78640   &   61600   &  38400    \\
10$^7$     &   100   &      &   x   &     5416   &   94584   &   18240  &   81760    \\
10$^7$     &   250   &      &   x   &     1169   &   98831   &    6189  &   93819    \\
\end{tabular}  
\label{table:stats}
\end{table*}

\subsubsection{Migration trap} \label{migration}

To quantify the rate of BHB hardening due to tertiary encounters at a migration trap in the AGN disk, we perform an additional set of Monte Carlo simulations.  We assume that the BHB orbits within a migration trap at distance $r_{\rm trap}$ from the central SMBH.  Although torques from gravitational interactions with the gas disk \citep[e.g.][]{Bartos17} are responsible for delivering the BHB to the migration trap (or its components, if the binary was formed there) and can even contribute to reducing its orbital separation, we ignore these effects here and focus on the hardening influence of gravitational scattering events with tertiary objects arriving at the trap.

All dynamical interactions are assumed to occur in the plane of the disk.  Hence, as before, we fix $n =$ 3 in Equation~(\ref{eqn:n}) and sample the eccentricities from Equation~(\ref{eqn:ecc2}).  For the relative velocity at infinity, we use Equation~(\ref{eqn:vrel2}) but replace the term $a_{\rm HS,disk}$ with the semi-major axis of the BHB $a_{\rm BHB}$ (hence, we ignore retrograde encounters, which will almost always be dissociative but are expected to be rare).  This is because we are only concerned with encounters for which significant energy is exchanged between the binary and single star and this only occurs if the encounter is direct (especially at high relative velocities).  As shown in Figure 3 of \citet{leigh16c}, the change in semi-major axis drops off dramatically beyond a distance of closest approach equal to the binary orbital separation.  The change in eccentricity drops off slightly more slowly, but is nonetheless negligible beyond a distance of closest approach a little more than twice the binary orbital separation.

We again assume a migration trap at $r_{\rm trap}$ $\sim$ 200$R_{\rm g}$ 
with no spherical component remaining at this radius (see Sect.~\ref{diskcomp}).
Assuming an SMBH mass $M_{\rm SMBH} =$ 10$^8$ M$_{\odot}$, a gas surface mass density $\Sigma =$ 10$^7$ M$_{\odot}$ pc$^{-2}$, 
Equation~(\ref{eqn:gamma_disk2}) yields $\sim$ 10$^3$ years for the mean time between encounters, assuming that the migrating objects are other black holes with $M_{\rm *} = 10 M_{\odot}$.  For migrating stars with $M_{\rm *} = 1 M_{\odot}$, this timescale is a factor of ten longer.  
Regardless, this is sufficiently short compared to the orbital period of the BHB at the location of the migration trap
of $\sim 10^3$ yr, 
so every encounter between the BHB and migrating disk stars should be direct, with impact parameter $\lesssim$ $a_{\rm BHB}$.  As already discussed, this estimate relies on a number of simplifying assumptions.  In real AGN disks, the true rate is likely to be time-dependent and depend sensitively on the evolution of the disk.  We are at the mercy of a number of uncertainties in the properties of these disks, due in large part to the lack of strong observational constraints for the properties of AGN disks, but also due to the analytic limitations of our model.  Nevertheless, the model shows that the encounter timescale can be very short relative to the case without a gaseous disk, due to the dissipative forces that it supplies.

\begin{figure}
\begin{center}
\includegraphics[width=\columnwidth]{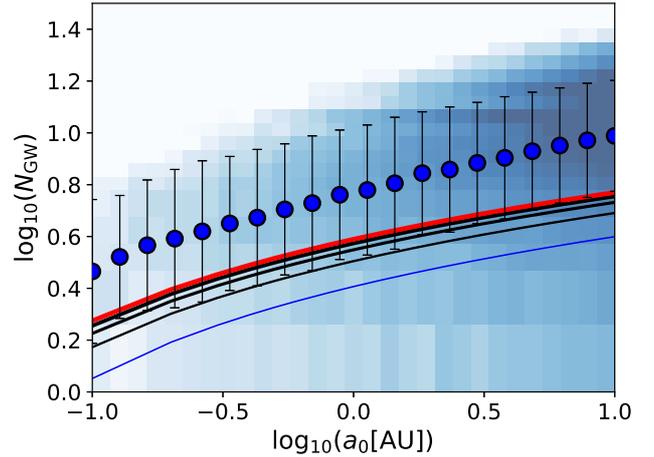}
\end{center}
\caption[The number of hardening interactions $N_{\rm GW}$ needed to harden a BHB to the point of rapid inspiral in an AGN disk without perturbations]{Same as Figure~\ref{fig:fig3} but for BHBs being hardened within the migration trap of an AGN disk.  We set $n =$ 3 in Equation~(\ref{eqn:n}) and hence consider only planar interactions.  The solid lines are the same as in Figure~\ref{fig:fig3} (i.e., they correspond to different values of $n$ in Equation~(\ref{eqn:n})), but assuming a relative velocity of zero.
\label{fig:fig7}}
\end{figure}

\begin{figure}
\begin{center}
\includegraphics[width=\columnwidth]{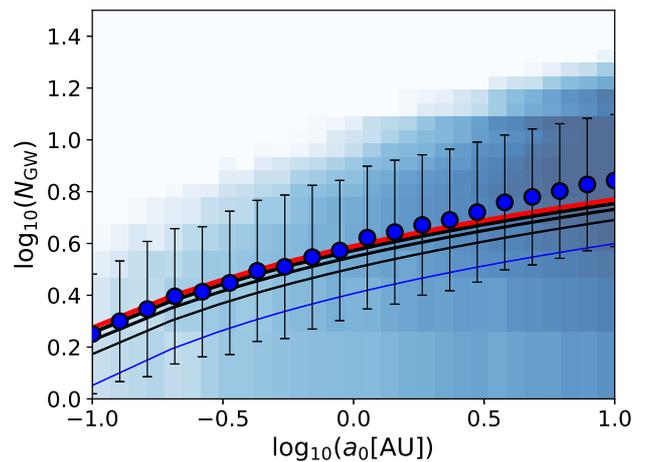}
\end{center}
\caption[The number of hardening interactions $N_{\rm GW}$ needed to harden a BHB to the point of rapid inspiral in an AGN disk with perturbations]{Same as Figure~\ref{fig:fig7} but including perturbing encounters.
\label{fig:fig8}}
\end{figure}

The results of this analysis are shown in Figures~\ref{fig:fig7} and~\ref{fig:fig8}.  As is clear, with or without perturbing encounters, the number of hardening interactions required before merger is always $\lesssim$ 10.  For a BHB in a migration trap, the results of our Monte Carlo simulations have the same dependence on orbital separation as our analytic predictions.  This is because the rate of encounters is sufficiently high that hardening due to GW emission is negligible, and dynamical interactions dominate the overall rate.  This is the case independent of the initial orbital separation of the BHB.  

We note an offset between our simulations and analytic predictions that is also seen in our previous comparisons, but was more difficult to quantify due to other effects also being significant.  Specifically, the simulations predict 2.5 times more hardening interactions $N_{\rm GW}$ to be required than the purely planar case with $n =$ 3.  Two possible explanations are, first, that we adopt the mode of the escaper velocity distribution in our derivation which does not always agree with the mean, since the distribution of escaper velocities is asymmetric.  Second, our Monte Carlo code outputs a discrete number of encounters that occurred before a given BHB merged, whereas our analytic formulation offers only a continuous function to describe the same number of hardening interactions.  This difference can become especially important when the number of hardening interactions is very low.  Nevertheless, these effects are minor and do not affect our overall results or conclusions.

As before, perturbing encounters have the effect of reducing the overall number of hardening interactions by a factor $\sim$ 2, by increasing the efficiency of GW emission in between successive direct encounters.  Finally, we note that, in the disk case, 100\% of BHBs merge in our Monte Carlo simulations, due to the low relative encounter velocities.

\begin{figure}
\begin{center}
\includegraphics[width=\columnwidth]{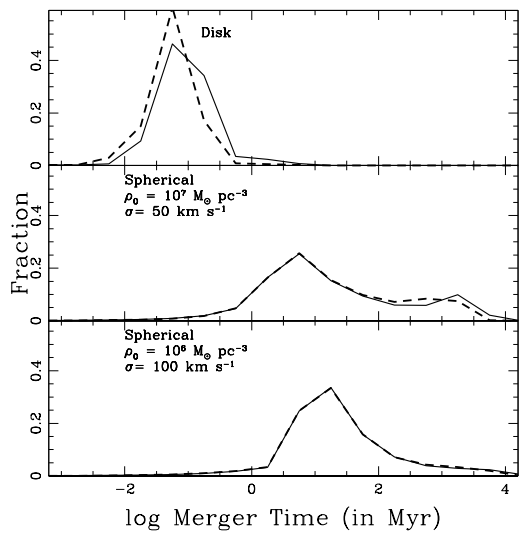}
\end{center}
\caption[The distributions of merger times for all realizations of our Monte Carlo simulations]{The distributions of merger times are summarized for all realizations of our Monte Carlo simulations, in both the spherical and disk components.  
     Note that merger times in disks are over an order of magnitude shorter than the spherical distributions.
Simulations without and with perturbations due to fly-by interactions are shown in 
  {\em dotted} and {\em solid} lines, 
respectively.  All merger times are given in megayears.
\label{fig:fig9}}
\end{figure}

In Figure~\ref{fig:fig9}, we show the distributions of merger times for all BHBs in our Monte Carlo simulations.  
Merger times are much less than a Hubble time for any 
    binary separation hard enough to avoid prompt disruption
The merger times in AGN disks are at least 
     an order of magnitude shorter than those in spherical distributions;
only about a factor of ten greater than the time-scale for direct encounters with single stars in the disk.  This motivates future studies of BHB mergers in AGN disks, to better understand contributing effects not considered in this paper.  For example, we have not accounted for any back-reaction of the BHB on the surrounding gas disk, which could open gaps in the disk if it is very thin and well-ordered \citep[e.g.][]{levin03,Paarde10}.  We have also neglected to account for a well-defined distribution of impact parameters between incoming single stars and the BHB.  Such effects should be properly studied and accounted for in future work.

\subsection{Dependence of mean merger time on v/$\sigma$} \label{rotation}

\begin{figure}
\begin{center}
\includegraphics[width=\columnwidth]{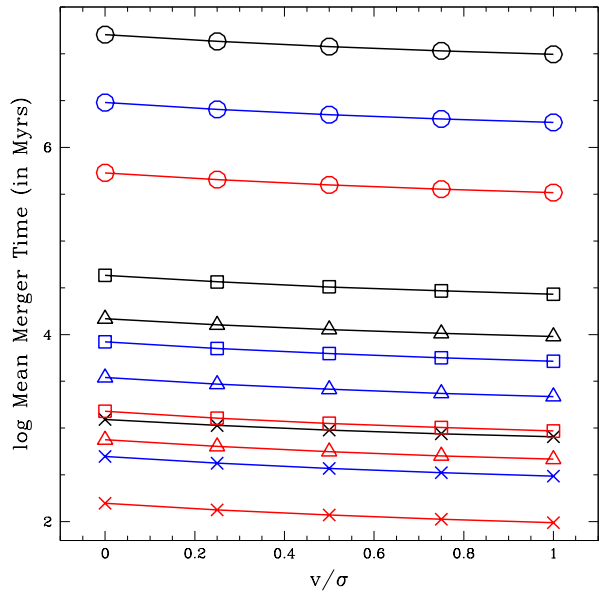}
\end{center}
\caption[The mean BHB merger times as a function of the ratio $v$/$\sigma$ for all realizations of our Monte Carlo simulations]{The mean BHB merger times are shown as a function of the ratio $v$/$\sigma$ for all realizations of our Monte Carlo simulations, as described in the text.  The circles, squares, triangles and crosses designate velocity dispersions of, respectively, $\sigma =$ 20, 50, 100 and 250 km s$^{-1}$.  The black, blue and red points designate central cluster densities of, respectively, $\rho =$ 10$^5$, 10$^6$ and 10$^7$ M$_{\odot}$ pc$^{-3}$.  All times shown are in Myr.  The solid colored lines connect each set of data points.
\label{fig:fig10}}
\end{figure}

Figure~\ref{fig:fig9} indirectly suggests that the BHB merger rate increases with increasing $v$/$\sigma$, where $v$ is the second-order velocity moment and $\sigma$ is the velocity dispersion.  We emphasize that here we discuss the ratio $v$/$\sigma$ in terms of flattened distributions (i.e., single-component 3-D clusters with anisotropic velocity distributions) and not disks, to illustrate that anisotropic velocity distributions could be more likely to host BHB mergers than purely isotropic distributions.  To quantify this effect, we run several additional realizations of our Monte Carlo simulations.  Adopting the same assumptions as before for the spherical case, we perform 10$^5$ realizations of our Monte Carlo simulations for every combination of velocity dispersion $\sigma =$ 20, 50, 100 and 250 km s$^{-1}$ and central cluster density $\rho =$ 10$^5$, 10$^6$ and 10$^7$ M$_{\odot}$ pc$^{-3}$.  We then calculate mean merger times for all BHBs that merge in our simulations, using the indicated value of v/$\sigma$ (meant as a proxy for the ratio of the second-order velocity moment and velocity dispersion for a given degree of anisotropy) to set the fractional contributions from planar and isotropic scatterings (e.g., $v$/$\sigma =$ 1 implies 50\% planar and 50\% isotropic scatterings).  The results of this exercise are shown in Figure~\ref{fig:fig10}.  

The message from Figure~\ref{fig:fig10} is clear:  the mean BHB merger time decreases only weakly with increasing $v$/$\sigma$.  A least-squares fit to these data suggests that the mean BHB merger time scales roughly as ($v/\sigma$)$^{-0.2}$.  This is independent of the host cluster density, velocity dispersion or the presence of perturbing encounters.  Thus, all other things being equal (e.g., the numbers of BHBs, their properties, etc.), \textit{we confirm that flattened distributions should host more BHB mergers, due to the increased probably of low-inclination (i.e., nearly planar) scatterings between the BHB orbital plane and interloping single stars.}  Importantly, Figure~\ref{fig:fig10} does not account for the higher incidence of dissociation due to encounters with cluster stars relative to disk stars, contributing to an even larger offset in the BHB merger rates in flattened versus spherical stellar systems relative to what is shown in Figure~\ref{fig:fig10}.  That is, none of the calculations shown in Table~\ref{table:stats} for the numbers of merged and disrupted BHBs in the cluster and disk components are considered in Figure~\ref{fig:fig10}.

Unlike in purely velocity-dispersion-supported clusters (such as globular clusters), a higher density does not necessarily translate directly in to a higher velocity dispersion in NSCs, but rather a higher root-mean-square velocity.  But, the root-mean-square velocity can be decomposed in to dispersion- and rotational-components.  Thus, for a given density, NSCs with a higher $v$/$\sigma$ should have higher rates of BHB mergers, both due to the lower relative velocities at impact during single-binary interactions, and the higher probability of having more (near-)planar interactions.

We emphasize that this result depends on the orientation of the BHB within the nuclear potential to be such that near-planar interactions occur more frequently than occurs in purely isotropic scatterings.  This is likely a sensitive function of the location of the BHB within the nucleus and the precise shape and properties of the gravitational potential.  In an AGN disk, for example, the BHB could experience Kozai-Lidov oscillations induced by the central SMBH.  Although the detailed evolution of this triple system depends sensitively on a number of different parameters \citep[e.g.][]{li14}, the BHB can be efficiently driven to merge by the quadrupole and octupole effects if the inclination angle falls in the nominal range (i.e., $\sim$ 50-90$^{\circ}$).  This could occur if some perturbing mechanism induces some small non-zero inclination of the BHB relative to the disk, which could then become amplified into the nominal Kozai-Lidov range via scattering interactions with tertiary objects in the disk.
A Kozai-Lidov-like effect could also arise within an axisymmetric potential if the BHB is located far from the origin, such that the BHB could be continually re-oriented within the host potential to favor planar interactions.  However, here it is unclear whether or not the time-scale could be sufficiently short to be relevant.  

\subsection{The fates of both the BHB and the scattered stars} \label{fate}

What will be the fates of those interloping single stars that are scattered off of the BHB?  Typical ejection velocities should be on the order of a few tens to $\sim$ 100 km s$^{-1}$.  Meanwhile, the escape velocity from such a deep potential well is $>$ 100 km s$^{-1}$ \citep{merritt04}.  Thus, the scattered stars will be ejected to larger radii within the disk (since planar scatterings eject the single star in the plane of the binary), and could eventually migrate back in toward the migration trap and scatter off the BHB a second time.  Interestingly, many single stars in the disk could form binaries, if their orbits about the central SMBH are packed sufficiently tight for the inter-particle distance to be comparable to the stars' Hill radii.  In this case, the binaries will migrate inward and interact with the BHB on a shorter timescale, resulting in a four-body scattering event.  This
problem is also analytically tractable using a similar methodology as used in this paper \citep{leigh16a}.  Again, however, this assumes strictly planar scatterings, and the overall picture will change as this assumption breaks down.  

Non-planar scattering interactions could act to increase the vertical component of the disk velocity dispersion and ultimately puff up the stellar component of the disk.  Simultaneously, they will increase the inclination of the BHB orbital plane relative to the plane of the disk.  If the timescale for non-planar scatterings to increase the inclination is longer than the Kozai-Lidov timescale at quadrupole order, then the eccentricity of the BHB could become very high due to Kozai-Lidov oscillations induced by the central SMBH \citep[e.g.][]{antonini16,leigh16a}. 
In this case, the BHB could merge due to GW emission on a short timescale.  

What will be the fate of a BHB in a migration trap \textit{after} it has merged?  If the spin angular momenta of the BHs are aligned then they will get only a small kick.  In this limit, the merged BHB could remain in (or quickly return to) the migration trap.  It is likely that the remaining BH will continue to interact with inward migrating stars and remnants in the disk.  The BH could grow steadily in mass (see \citealt{horn12} for the analogous situation in a protoplanetary disk), either by accreting from the surrounding gaseous disk, merging with incoming stars in the disk (either directly or by first forming a binary that gets rapidly ground down to merger via dynamical scatterings), or eventually forming new BHBs that subsequently merge via dynamical scattering and GW emission.  Indeed, it is possible that sufficient mass growth could occur that an intermediate-mass black hole (IMBH) forms in the migration trap.  This offers a potential channel for large mass ratio inspiral events, if additional stellar-mass BHs in the disk become bound to the IMBH and merge with it.

\section{Summary and Discussion} \label{discussion}

     We find
two separate environmental extremes in which a BHB in a galactic nucleus can be efficiently hardened via three-body scattering interactions to the point of rapid inspiral due to GW emission.  These are: (1) spheroidal clusters with high densities, low to moderate velocity dispersions and no significant Keplerian component; and (2) migration traps in disks around massive SMBHs but without any significant spherical stellar component in the vicinity of the migration trap, perhaps because of orbital inclination reduction by the disk.  Importantly,
    both of these cases works for old stellar populations; neither requires 
recent star formation or globular cluster infall,
although these events can certainly deliver new BHBs to the nucleus, and replenish an otherwise dynamically depleted population.  We argue that 
    such replenishment is only necessary
to observe 
BHB mergers in nuclei that do not meet criteria (1) or (2) described above.

We 
     derived
an analytic formalism for 
     the evolution 
of the binary orbital parameters of a stellar-mass BHB being hardened in a dense galactic nucleus.  Specifically, we calculate the number of interactions $N_{\rm GW}$ needed to harden a stellar-mass BHB to the point that the timescale for inspiral due to gravitational wave emission is shorter than the time for a subsequent three-body scattering event (i.e.\ $\tau_{\rm GW} < \tau_{\rm 1+2}$, where $\tau_{\rm 1+2} \sim$ 1--10 Myr).  This is to assess the contribution of dynamical hardening to the rate of BHB mergers in galactic nuclei.  
     We then made a Monte Carlo calculation of
$N_{\rm GW}$ to include the loss of binary orbital energy and angular momentum due to GW emission that occurs in between scattering events.  This allows us to quantify the relative contributions to the BHB merger rate from dynamical hardening and GW emission.  We explore a range of nuclear environments, including migration traps.

     The Monte Carlo calculation shows
 that, at the influence radius of a Milky Way-like NSC with $M_{\rm SMBH} =$ 10$^6$ M$_{\odot}$ and $\sigma =$ 100 km s$^{-1}$, hardening and circularization due to GW emission that occurs between single-binary encounters can reduce 
     $N_{GW}$ 
by a factor of order unity.
As the BHB migrates inwards toward the central SMBH, the encounter rate 
   increases 
(assuming gravitational focusing is negligible) and the total number of scattering events $N_{\rm GW}$ 
    approaches the analytic estimate 
(Eq.~\ref{eqn:numhard3}) as the effects of inter-encounter GW emission are reduced due to the higher encounter frequency.  With that said, the relative velocity at infinity can also increase significantly, increasing the probability that the BHB will be dissociated during an energetic single-binary interaction before it can merge due to GW emission.  


In a migration trap,
hardening due to dynamical interactions can be 
efficient.  
    The disk ensures that objects on prograde orbits that enter the trap will engage in planar 
    encounters at low relative velocities, ensuring hardening rather than dissociating encounters.
    The
rate of encounters is sufficiently high that hardening due to GW emission is negligible, so dynamical interactions dominate the overall rate,
   and $N_{GW}$ agrees well with our analytic estimate.
     The
tertiary scattering rate for BHBs in migration traps adopted in this paper
    could
increase by three orders of magnitude without affecting the conclusion that dynamical interactions dominate the rate of binary hardening.  
With that said, however, we 
    do
emphasize that there is uncertainty in our calculated rates for the AGN disk model.  For example, our uncertainties in the disk aspect ratio
   alone can introduce four orders of magnitude of uncertainty to the encounter rate in the disk
(see Section~\ref{diskcomp}).  
    Nevertheless, this timescale is so
short, relative to the timescale for encounters with single stars in the absence of the dissipative forces supplied by a gaseous disk,
    that it is likely to remain shorter regardless of the disk structure.  Interestingly, our results suggest that this time-scale could be even shorter than the main-sequence lifetimes of massive stars, such that the stars merge before becoming BHs.  This would not significantly change our conclusions, however, but instead accelerate the formation of massive BHs in AGN disks, since more massive BHs should migrate faster.

In a migration trap, the Keplerian shear could terminate scattering interactions artificially early relative to the equivalent scattering interaction in isolation if the interloping single star wanders beyond the Hill radius before becoming completely unbound from the BHB.  However, in this environment, the Hill radius of a 
BHB can be comparable to its orbital separation, and most three-body scatterings should be prompt \citep[e.g.][]{leigh16c,leigh16d}.  Only a few temporary ejection events where the single star experiences a prolonged excursion but remains bound to the BHB, should occur before the interloping single star escapes completely, with the typical excursion distance being comparable to the BHB orbital separation.  Therefore, we expect relatively few three-body scattering interactions to be terminated early.  Given that the initial relative velocity between the BHB and the interloping star is very small, few if any of these interrupted encounters should soften the BHB.  Nonetheless, the single star will ultimately remove less energy from the BHB's orbit when the interaction is terminated artificially.  This could increase the number of scattering interactions required to harden the BHB to merger, relative to our results shown in Figures~\ref{fig:fig7} and~\ref{fig:fig8}.

We further find a smaller effect: higher rates of BHB mergers in NSCs with higher specific angular momentum (i.e., a higher value for the ratio $v$/$\sigma$).  That is, to first-order, rotation-supported NSCs are more conducive to BHB mergers than are pressure-supported NSCs (assuming planar and isotropic scatterings are, respectively, more common in the former and latter nuclear environments).  There are two reasons for this.  First, as shown in Figure~\ref{fig:fig10}, the mean BHB merger time scales decreases slightly with increasing $v$/$\sigma$; it falls off roughly as ($v/\sigma$)$^{-0.2}$. Second, the encounter energy is independent of orientation, and planar scatterings are more effective at hardening BHBs than are isotropic scatterings.  Hence, as shown in Table~\ref{table:stats}, it is less likely that a BHB is dissociated during a second encounter in the planar case, relative to the isotropic case.  This leads to more disrupted binaries in the isotropic case, and predicts a higher rate of BHB mergers in NSCs with higher ratios $v$/$\sigma$, assuming all else is the same, including the numbers of BHBs, their orbital parameters, the encounter rate, the age of the stellar population(s), etc.  Interestingly, \citet{petrovich17} also found that non-spherical clusters can enhance the rates of BH-BH binary mergers relative to spherical clusters, but in their case the underlying mechanisms responsible are cluster-enhanced Lidov-Kozai oscillations induced by the central SMBH and chaotic diffusion of the orbital eccentricities due to resonance overlap.

The overall result of this study is clearly shown in Figure~\ref{fig:fig9}: planar scattering in disks results in merger times an order of magnitude faster than in even dense spherical systems.  As a result, tertiary encounters between stars and BHBs in nuclear disks could contribute significantly to the rates of BHB mergers detected by aLIGO.  We have made a number of simplifying assumptions, but do not believe they will change the qualitative difference between the two mechanisms.  The detailed balance between them will require more sophisticated modeling.  The treatment applied here does neglect the detailed dynamics of a cluster with significant rotation.  In particular, the rotation- and dispersion-dominated components are free to interact in potentially complicated ways not accounted for here.  In principle, a Fokker-Planck model can be applied to properly calculate the interaction rates, while simultaneously accounting for the orientation of the BHB orbital plane relative to the host NSC potential.  Unfortunately, however, observational constraints for the detailed kinematic behaviour of galactic nuclei are lacking.  Consequently, it is not always clear or straight-forward how to choose a representative distribution function for the stellar population in a given nucleus, and simplifying assumptions must always be made.  

\section{Acknowledgments}

The authors kindly thank an anonymous reviewer for 
useful comments
    that
improved the manuscript.  NWCL acknowledges support from a Kalbfleisch Fellowship at the American Museum of Natural History.  BM \& KESF are supported by NSF PAARE AST-1153335 and NSF PHY11-25915.  M-MML and NWCL were partly supported by NSF AST11-09395.  ZH is partly supported by NASA ATP grant NNX15AB19G
   and
a Simons Fellowship for Theoretical Physics.

\end{document}